\pdfoutput=1
\documentclass[sigconf,natbib=false]{acmart}
\usepackage{mathtools,nicefrac}
\usepackage{csquotes}
\usepackage[backend=biber,style=ieee,sortlocale=en-US,url=true,doi=false,date=year,minnames=2,maxnames=4,isbn=false, sortcites, dashed=false]{biblatex}
\usepackage[textsize=tiny]{todonotes}
\usepackage{tikz}
\usepackage[binary-units=true]{siunitx}
\usepackage{booktabs}
\usepackage{hyperref}

\DeclareMathOperator{\Tr}{Tr}

\newcommand{\bra}[1] {\ensuremath{\langle #1 |}}
\newcommand{\ket}[1] {\ensuremath{| #1 \rangle}}
\newcommand*{\qop}[1]{\ensuremath{\mathit{#1}}}
\newcommand{\cnot}{\qop{CNOT}}
\newcommand{\cex}{\qop{CEX}}
\newcommand{\csum}{\qop{CSUM}}

\usetikzlibrary{backgrounds,decorations.pathmorphing,decorations.pathreplacing,positioning,matrix}

\hypersetup{
	pdftitle={Compilation of Entangling Gates for High-Dimensional Quantum Systems},
	pdfsubject={28th Asia and South Pacific Design Automation Conference (ASP-DAC) 2023},
	pdfauthor={Kevin Mato, Martin Ringbauer, Stefan Hillmich, and Robert Wille} %
}

\setcopyright{none}
\copyrightyear{2023}
\acmYear{2023}
\acmDOI{10.1145/3566097.3567930}
\acmConference[ASP-DAC '23]{Asia and South Pacific Design Automation Conference}{January 16--19, 2023}{Tokyo, Japan}

\let\origtheequation\theequation
\makeatletter
\def\tagform@#1{\maketag@@@{\ignorespaces#1\unskip\@@italiccorr}}
\makeatother
\renewcommand{\theequation}{(\origtheequation)}

\AtBeginDocument{%
}
\begin{document}

\title[Compilation of Entangling Gates for High-Dimensional Quantum Systems]{Compilation of Entangling Gates\\for High-Dimensional Quantum Systems}

\author[Kevin Mato, Martin Ringbauer, Stefan Hillmich, and Robert Wille]{Kevin Mato$^*$\hspace{3.0em} Martin Ringbauer$^+$\hspace{3.0em}Stefan Hillmich$^{\dagger}\hspace{3.0em}$Robert Wille$^{*\ddag}$\hspace{3.0em}}
\affiliation{%
   \institution{\vspace{6pt}$^*$ Chair for Design Automation, Technical University of Munich, Germany}
}
\affiliation{%
  \institution{$^+$Institute for Experimental Physics, University of Innsbruck, Austria}
}
\affiliation{%
  \institution{$^\dagger$Institute for Integrated Circuits, Johannes Kepler University Linz, Austria}
}
\affiliation{%
  \institution{$^\ddag$ Competence Center Hagenberg (SCCH) GmbH, Austria}
}
\email{kevin.mato@tum.de , martin.ringbauer@uibk.ac.at, stefan.hillmich@jku.at, robert.wille@tum.de}
\email{https://www.cda.cit.tum.de/research/quantum/}

\begin{abstract}
    Most quantum computing architectures to date natively support multi-valued logic, albeit being typically operated in a binary fashion. 
    Multi-valued, or qudit, quantum processors have access to much richer forms of quantum entanglement, which promise to significantly boost the performance and usefulness of quantum devices. 
    However, much of the theory as well as corresponding design methods required for exploiting such hardware remain insufficient and generalizations from qubits are not straightforward.
    A particular challenge is the compilation of quantum circuits into sets of native qudit gates supported by state-of-the-art quantum hardware.
    In this work, we address this challenge by introducing a complete workflow for compiling any two-qudit unitary into an arbitrary native gate set. 
    Case studies demonstrate the feasibility of both, the proposed approach as well as the corresponding implementation (which is freely available at \href{https://github.com/cda-tum/qudit-entanglement-compilation}{\emph{github.com/cda-tum/qudit-entanglement-compilation}}).
\end{abstract}


\maketitle

\section{Introduction}

In the future, quantum computers are expected to solve industrial and scientific problems with a reduced consumption of resources and greater algorithmic efficiency. 
Current quantum devices can host up to hundreds of noisy \emph{quantum bits}~(qubits) and support a limited number of logical operations on these qubits. For this reason, they are referred to as \emph{Noisy Intermediate-Scale Quantum}~(NISQ) devices~\cite{preskill2018quantum}.
A variety of technology platforms have now implemented such NISQ devices, including superconducting circuits~\cite{Arute2019}, trapped ions~\cite{Pogorelov2021}, and single photons~\cite{Flamini2018a}. 
Notably, while these devices thus far almost exclusively work with two-level qubits, the underlying hardware almost always natively supports encoding multi-valued logic in high-dimensional \emph{quantum digits}~(qudits).

The research on qudit design and computation has a long history, with efforts primarily focusing on conceptual studies of algorithms for idealized qudits and their comparison to qubits~\cite{Wang2020}. Fundamentally, a qudit can not only store and process more information per quantum particle, but also features a richer set of logical operations~\cite{Huber2013} that make processing more efficient. 
\mbox{Proof-of-principle} demonstrations~\cite{ringbauer2021universal,Lanyon2008,Gedik2015,Zhan2015} have shown that qudits enable improvements in circuit complexity and algorithmic efficiency for a wide class of problems. 
These results inspired proposals for and demonstration of qudit basic control in numerous physical platforms such as trapped ions~\cite{Zhang2013Contextuality,ringbauer2021universal}, photonic systems~\mbox{\cite{Lanyon2008,Ringbauer2017Coherence, Hu2018a, Malik2016}}, superconducting circuits~\cite{Kononenko2020,Morvan2020}, Rydberg atoms~\cite{Ahn2000, Weggemans2021}, nuclear spins~\cite{Godfrin2017}, cold atoms~\cite{Anderson2015, Kasper2020}, and nuclear magnetic resonance systems~\cite{Gedik2015}. 
More recently, efforts have intensified with the demonstration of universal qudit quantum processors with error rates that are competitive to qubit systems~\cite{ringbauer2021universal,Chi2022}.

However, a key task for using qudit quantum processors remains the efficient compilation of circuits for this hardware. 
Compared to qubits, where every entangling gate is equivalent to the \qop{CNOT}~\cite{NielsenChuang} gate, qudits offer much richer possibilities in the form of many inequivalent entangling~gates. 
The flip side of these opportunities, however, are challenges in finding suitable gate sets that are native to the hardware and, then, compiling algorithms to these gate sets. 
Thus far, much of the theory required to address these challenges is insufficient and, accordingly, no corresponding design methods are available for this purpose yet---making the compilation of entangling gates a mostly manual task and, thus, preventing the further exploitation of high-dimensional quantum systems.

In this work, we introduce a workflow for the compilation of arbitrary two-qudit entangling gates in any high-dimensional system.
The core idea is to map the target unitary operating on two qudits to a single-qudit unitary in an appropriately larger space.
The latter can then be decomposed into two-level couplings using established techniques~\cite{Bullock_2005}.
The resulting decomposition will feature at most $d^2$ two-level entangling gates of two standardized types. 
For decomposing these two standard gates into any \mbox{hardware-native} gate set, we developed an offline optimization routine, which we demonstrate on a recently developed gate set for trapped ion qudit quantum processors~\cite{ringbauer2021universal,nativequdit}.
In typical experimental scenarios, where the native gates act only on a subspace of the full Hilbert space, the cost of this pre-computation is independent of the size of the target unitary. 
The resulting circuit will express the target unitary using only native gates. 
Overall, this results in a methodology that, for the first time, enables compiling entangling gates for high-dimensional quantum systems in a fully automatic fashion. 
The proposed method can be applied to any two-qudit unitary and is computationally efficient, since the numerical optimization step is done offline.
Case studies demonstrate the feasibility of the proposed method and a corresponding implementation is available in open-source via \href{https://github.com/cda-tum/qudit-entanglement-compilation}{\emph{github.com/cda-tum/qudit-entanglement-compilation}}, along with other components of the Munich Quantum Toolkit (MQT).
The proposed tool is only a component of a complete compiler for high-dimensional and mixed-dimensional systems, that realizes a fully automated and computationally scalable workflow for the compilation of entangling operations.
Therefore the results proposed are not necessarily efficient to the point of practical applicability on a quantum system, both in terms of gate count and possible requirement of additional circuit optimization.

The remainder of this paper is structured as follows: 
\autoref{sec:background} briefly reviews the basics of quantum information processing (QIP) and entanglement. 
\autoref{sec:motivation} motivates the problem of entanglement compilation for qudits. 
\autoref{sec:entanglement-compilation} describes the proposed approach to tackle entanglement compilation. 
\autoref{sec:case-studies} provides case studies on the feasibility of the proposed approach. 
Finally, \autoref{sec:conclusion} concludes the paper.

\section{Background}
\label{sec:background}
In this work, we provide the basis for efficient compilation of entangling operations for high-dimensional systems. To this end, this section first briefly reviews the fundamentals of QIP (with a focus on high-dimensional quantum logic) and entanglement. %

\subsection{Quantum Information Processing}
The fundamental unit of classical information is the \emph{bit} (binary digit), which can exclusively take the values 0 or 1. Generalizing this concept to quantum computers gives rise to the \emph{qubit} as the corresponding unit of quantum information. The crucial difference to the classical case, however, is that qubits can be in any linear combination of the basis states $\ket{0}$ and $\ket{1}$ (using Dirac's bra-ket notation). 

Since there are no ideal two-level systems in nature, qubits are usually constructed by restricting the natural multi-level structure of the underlying physical carriers of quantum information. These systems, therefore, natively support \mbox{multi-level logic} with the fundamental unit of information termed a \emph{qudit}.
A qudit is the quantum equivalent of a $d$-ary digit with $d\geq 2$, whose state can be described as a vector in a $d$-dimensional complex Hilbert space $\mathcal{H}_d$. 
The quantum state $\ket{\psi}$ of a qudit can thus be written as a linear combination
$
    \ket{\psi} = \alpha_0 \cdot \ket{0} + \alpha_1 \cdot \ket{1} + \ldots + \alpha_{d-1} \cdot \ket{d-1},
$
or simplified as vector $ \ket{\psi} = \begin{bsmallmatrix} \alpha_0 & \alpha_1 & \ldots & \alpha_{d-1}\end{bsmallmatrix}^\mathrm{T}$, where $\alpha_i \in \mathbb{C}$ are the amplitudes relative to the computational basis of the Hilbert space---given by the vectors $\ket{0}, \ket{1},\ket{2},\ldots, \ket{d-1}$.
The squared magnitude of an amplitude $|\alpha_i|^2 $ gives the probability with which the corresponding basis state $i$ would be observed when measuring the qudit in the computational basis. Normalization of probabilities further requires that $\sum_{i=0}^{d-1} |\alpha_i|^2 = 1$. %
\begin{example}\label{ex:state}

    Consider a single qudit with three levels (also referred to as \emph{qutrit}).
    
    The quantum state 
    \(\ket{\psi} = \sqrt{\nicefrac{1}{3}}\cdot\ket{0} + \sqrt{\nicefrac{1}{3}}\cdot\ket{1} + \sqrt{\nicefrac{1}{3}}\cdot\ket{2}\)
    is a state with equal probability of measuring each basis. 
    A different notation for the same quantum state may be  \mbox{$\sqrt{\nicefrac{1}{3}}\cdot \begin{bsmallmatrix} 1 & 1 & 1\end{bsmallmatrix}^\mathrm{T}$}.
\end{example}

Two key properties that distinguish quantum computing from classical computing are superposition and entanglement.
A qudit is said to be in a \emph{superposition} of states in a given basis when at least two amplitudes are non-zero relative to this basis. 
\emph{Entanglement}, 
instead, describes a powerful form of non-classical correlation born from interactions in multi-qudit systems. 
Quantum computer operations
are represented by unitary matrices $U$, satisfying $U^\dagger U = U U^\dagger = I$.

\subsection{Entanglement Structures}
Classically, the state of a bipartite system can always be written as a product of the state of the individual systems (or a mixture of such products).
Entangled quantum states encode information in a non-local way, such that it can only be extracted from the full system, but not from the constituent qudits.
More precisely, two or more quantum systems are in an \emph{entangled} state, whenever their state cannot be written as a product of states of the individual subsystems (or a convex mixture of such products)~\cite{NielsenChuang}. While systems of two qubits are quite well understood, entanglement in multi-partite or higher dimensional systems still present a range of open questions~\cite{horodecki2009quantum}.

The central scope of the work are quantum logic operations generating quantum entanglement, with a special focus on bipartite entangling gates, which enable the core aspect of error correction. Generalizations to multipartite entangling operations will be tackled where appropriate. The first notable observation is the much richer entanglement structure of qudits compared to qubits.

Specifically, for qubits, all entangling operations are related to the controlled-\qop{NOT} ($\cnot$) gate via local operations on the subsystems~\cite{NielsenChuang}. Hence, for qubit systems, it is sufficient to compile the $\cnot$ gate to the native operations of the quantum hardware. For qudit systems, instead, this is no longer true and they can be entangled in many inequivalent ways. Consequently, while any single entangling gate is sufficient for universal quantum computation~\cite{brennen2005efficient}, not all entangling gates are equally useful for any given application.

\begin{example}\label{ex:entangling-gates}
Consider, the controlled-exchange gate $\cex$~\cite{ringbauer2021universal} defined by
\begin{align}
    \cex_{c,t_1,t_2} : \begin{cases}
    \ket{c,t_1} \leftrightarrow \ket{c,t_2} \\
    \ket{j,k} \rightarrow \ket{j,k}  \quad\text{for $j\neq c, k\neq t_1,t_2$} .
    \end{cases}
    \label{eq:cex}
\end{align}
This qudit-embedded version of the $\cnot$ gate generates qubit-level entanglement in a high-dimensional Hilbert space. However, there are gates that directly generate \emph{genuine} qudit entanglement, such as the controlled-SUM gate defined by
\begin{align}
    \csum: \ket{i,j} \mapsto \ket{i,i\oplus j} ,
    \label{eq:csum}
\end{align}
where $\oplus$ denotes addition modulo $d$. These are just two examples of a more general theme in qudit systems, where entangling gates differ in their \emph{entangling power} or \emph{structure}, that corresponds to the amount of entanglement generated by an operation. The reasons why genuine qudit gates produce more entanglement than the first one are more intuitively explained in Ref.~\cite{nativequdit}. While the $\cex$ gate has a low entangling power, since it only generates two-level entanglement, it is very flexible and can be used to intuitively build up any entanglement structure. 
In turn, however, it is quite inefficient for constructing highly entangling unitaries such as the $\csum$ gate out of two-level entangling operations. 
Too much entangling power, however, is not always helpful either. It may happen that a qudit circuit requires the generation of entanglement only within a small subspace of the Hilbert space, in that case applying $\csum$ becomes disadvantageous and the $\cex$ is preferable.

\end{example}

In practice, of course, neither of the gates from \autoref{ex:entangling-gates} might be natively supported by the quantum hardware. 
However, potent qudit quantum devices are expected to 
support several primitive entangling gates,
with a range of entangling powers. 
Hence, 
it becomes crucial to be able to compile arbitrary qudit unitaries into the native set of operations, while making the best use of the available resources.

\section{Motivation}
\label{sec:motivation}
Based on the general setting introduced above, the compilation of entangling operations into native gate sets is of central importance for efficient QIP. Since the general problem is NP-hard already for qubits~\cite{Botea2018OnTC}, quantum computers critically depend on efficient, if not optimal, compilation methods. Given the more complicated entanglement structure of qudits, compared to their binary counterpart, the compilation problem becomes much more challenging in high-dimensional spaces.

\subsection{Problem Formulation}
Qubit circuits may be written using several entangling gates, all equivalent to the $\cnot$ gate~\cite{bullock_23_elementary}.
For qudit systems, the situation is very different. With the structure of entanglement becoming much richer, there is not just more freedom in designing quantum circuits, but also more opportunities for optimizing their efficiency with the right set of operations. The challenge for qudit compilation is then to understand and to harness this entanglement structure for efficient decompositions while at the same time, keeping the problem computationally tractable.

\begin{example}

    In order to make the problem more intuitive for the qubit expert, the use of a metaphor can help. Using color coding as an analogy: Information encoded in one qubit is like a single grayscale pixel. Information encoded in a \emph{qutrit} ($d=3$) is like an RGB pixel that can combine 3 different colors and all their possible shades inbetween. 
    Now, adding a second pixel, the qubits will only have two parameters, where qutrits have 6 parameters for leveraging the full potential of the two pixels.
    This highlights the drastically higher information density in qutrits (and subsequently qudits of even higher dimensions).
    However, the qutrit, just like the color pixel, requires a vastly more complex control system, to efficiently use the capability.

\end{example}

The problem of entanglement compilation can now be formulated as follows: 
\emph{Given a unitary~$U$ representing an interaction between two qudits of dimension $d$, find a decomposition of~$U$ into arbitrary local unitaries and a pre-defined set of entangling gates, in a way that is as close to the optimum as possible.}
In an application setting, the native gate set will be defined by the used quantum hardware, which will also set the cost of each component of the decomposition. While it is in general not known how complex the optimal solution is, the general figure of merit will be the gate fidelity given the experimental noise sources for the various gates. Here, we will focus on two-qudit gates, although the methods we present generally also apply to the multi-qudit setting, at the cost of decomposing the tensor product in two-qudit unitaries.

\subsection{State of the Art}
In this section, we briefly review existing approaches to compile entangling operations in high-dimensional Hilbert space.
While some results exist for special cases~\cite{smith2019entanglement,Bullock_2005}, these are of limited use in actual quantum hardware, which typically does not natively support the types of interactions we might want to work with.

More precisely, in pioneering work, Ref.~\cite{Bullock_2005} introduced a synthesis algorithm for qudit entangling gates that produces decompositions in terms of controlled-householder rotations. They prove a lower and a constructive upper bound on the number of two-qudit \mbox{controlled-rotation} gates required for an arbitrary two-qudit unitary. These results, however, only apply to the specific controlled rotations used, without considering the constraints or indeed opportunities of physical qudit quantum processors.
In Ref.~\cite{Heyfron2019AQC}, a compilation algorithm for generic qudit unitaries is presented. The final goal of the mathematical procedure is to reduce the number of gates that use magic state injection protocols. The entanglement complexity of qudit systems is ignored, and although an elegant solution is found, the restricted scalability of the algorithm and the final output, too far from being applied to a physical quantum processor, make the work still theoretical.
Other previous works~\cite{glaudell2019quantum} try to lay out routines for the synthesis of qudit circuits, in this particular case for qutrits only.

Related to this, Ref.~\cite{smith2019entanglement} presents a method to construct circuits for generating different forms of qudit entanglement using a specific gate set including the controlled-exchange gate. While the goal is the generation of specific quantum states from a fixed input, it is unclear to what extent these methods can be transferred to the more complicated compilation of entangling unitaries.

In Ref.~\cite{Martinez_2016}, the challenge of multi-qubit compilation is tackled. The authors use parametrized quantum circuits and numerical optimization of the fidelity function, for solving an incremental structure of alternating local and entangling gate layers. More precisely, the work focuses on the compilation of global entangling gates based on an ansatz made of alternating global entangling gates and specific equatorial rotations; the approach cannot be directly applied to qudit entanglement compilation, due to the inefficient search strategy. Moreover, the solution provided in~\cite{Martinez_2016} is unable to express the richer variety of local and partial entangling operations available in this new setting.
Finally, in Ref.~\cite{Friis2014} arbitrary controlled unitaries are proved to be physically implemented using auxiliary levels in the qudit Hilbert space, regardless of their form.

\section{Efficient Compilation of Entanglement}
\label{sec:entanglement-compilation}
Motivated by the challenges described in \autoref{sec:motivation}, we now propose a workflow that enables the compilation of any two-qudit unitary into any target gate set. The proposed solution consists of two steps. First, a synthesis method, which takes the target two-qudit unitary and returns a high-level circuit comprised of only two types of \mbox{two-level} entangling gates, regardless of the initial unitary. This step is crucial for making the technique scalable to arbitrary dimensions. 
Second, a compilation step that uses parametrized circuits and numerical optimization to decompose the two standardized two-level entangling gates into any target gate set. Although this step can be computationally intensive, it can be pre-computed, since the structure from the first step is fixed. Hence, this step does not affect the scalability of the method.

\subsection{ Step 1: QR Decomposition of Entangling Operations}

Qudit quantum hardware typically exploits two-level couplings between the various qudit levels, despite having access to a full high-dimensional Hilbert space. It has been shown that this is sufficient for implementing arbitrary single-qudit operations with a cost that is at most quadratic in the size of the Hilbert space~\cite{Bullock_2005}. In the simplest instance, the resulting sequence of operations is composed of two-level \emph{Givens}-rotations between adjacent sites $V_{i}$ and a diagonal phase matrix $\Theta$ in $U = V_k\cdot V_{k-1}\cdot\ldots\cdot V_1\cdot\Theta$~\cite{Bullock_2005}.
This algorithm is universally valid.
However, it scales quadratically in the Hilbert space dimension and, due to the rigid structure, it can introduce redundant operations. Hence, for efficiently compiling local qudit unitaries, more advanced algorithms, that take into account the structure of the qudits and experimental constraints, can be beneficial. For the purpose of the present work, however, it is precisely the standardized structure of the QR decomposition that is beneficial.

In more detail, we start by interpreting the target two-qudit unitary $U$ as a single qudit unitary in dimension $d^2$. This is achieved by mapping the two-qudit states to a single qudit as $\ket{i,j} \mapsto \ket{d\cdot i+j}$. We then apply the QR decomposition algorithm to that \mbox{higher-dimensional} local unitary. Without loss of generality, we thereby assume a ladder-type coupling, where each of the virtual \mbox{single-qudit} states is coupled to its neighboring states $\ket{j}\leftrightarrow\ket{j\pm 1}$. 
The QR~decomposition then applies successive rotations between adjacent state in the virtual single qudit. 
The result of the decomposition is a set of two-level rotations of the form 
\begin{align*}
    R(\theta,\phi) = 
    \begin{bmatrix}
        \cos\frac{\theta}{2}                        & \sin\frac{\theta}{2} (-i\cos\phi - \sin\phi)  \\
        \sin\frac{\theta}{2}(-i\cos\phi + \sin\phi) & \cos\frac{\theta}{2} \\
    \end{bmatrix}
\end{align*}
followed by a phase matrix, which can be represented as a sequence of phase rotations on neighboring states, which again can be decomposed into two-level rotations using the identity $\qop{Z}(\theta)= R(\tfrac{\pi}{2}, 0)\cdot R(\theta,\tfrac{\pi}{2} )\cdot R(-\tfrac{\pi}{2}, 0)$.

As a consequence of our choice of ladder-type coupling, the two-level rotations all occur between adjacent states and are thus embedded into the $d^2$ dimensional Hilbert space as 
\begin{align*}
    \hat{R}_{i}(\theta,\phi) = 
    \begin{bmatrix}
        1 & \cdots & 0 \\
        0 & \cdots [R(\theta,\phi)]_{i,i+1} \cdots & 0 \\
        0 & \cdots & 1\\
    \end{bmatrix} ,
\end{align*}
where the subscript ${i,i+1}$ denotes the affected states. 
The resulting matrix is a two-level rotation \qop{\hat{R}} embedded in a higher-dimensional space that translates to an entangling operation in the original two-qudit system.
The next challenge is now decomposing these entangling gates into the native gate set of the target hardware. 

A straightforward approach would be to simply try and compile each of the $\mathcal{O}(d^2)$ entangling gates in the sequence into the target gate set, although computationally very costly. Instead, one can notice that only two different gates need to be compiled.
This is most easily seen at a simple example with two qutrits. By design, each of the two-level rotation $R_i$ 
is represented as $2\times 2$ block matrices in a $d^2 \times d^2$ identity matrix as in \autoref{eq:cRpS}.
\begin{align}
    \scalebox{0.70}{\mbox{\ensuremath{\left[ \begin{array}{c c c|c c c|c c c}
       \cdot & \cdot  & \cdot & \cdot & \cdot  & \cdot & \cdot & \cdot  & \cdot \\
       \cdot & cR  & cR & \cdot & \cdot  & \cdot & \cdot & \cdot  & \cdot \\
       \cdot & cR  & cR & \cdot & \cdot  & \cdot & \cdot & \cdot  & \cdot \\
      \midrule
       \cdot & \cdot  & \cdot & \cdot & \cdot  & \cdot & \cdot & \cdot  & \cdot \\
       \cdot & \cdot  & \cdot & \cdot & \cdot  & \cdot & \cdot & \cdot  & \cdot \\
       \cdot & \cdot  & \cdot & \cdot & \cdot  & pS & pS & \cdot  & \cdot \\
      \midrule
       \cdot & \cdot  & \cdot & \cdot & \cdot  & pS & pS & \cdot  & \cdot \\
       \cdot & \cdot  & \cdot & \cdot & \cdot  & \cdot & \cdot & \cdot  & \cdot \\
       \cdot & \cdot  & \cdot & \cdot & \cdot  & \cdot & \cdot & \cdot  & \cdot
    \end{array}\right]}}}
    \label{eq:cRpS}
\end{align}
Here, the horizontal and vertical lines indicate the tensor product structure of the original two-qutrit unitaries. We note that whenever the rotation $R_i$ acts on levels that do not cross these lines, it physically corresponds to a controlled-rotation gate \qop{cRot} in the two-qudit system (indicated above as \qop{cR}). When $R_i$ does cross a boundary, it corresponds to a partial Swap operation \qop{pSWAP} in the bipartite system (indicated above as \qop{pS}). While mathematically very similar, these operations are fundamentally different in their complexity, so they must be treated separately. 
Fortunately, it is sufficient to compile just one example of each, since the other cases (for different qudit indices) can be obtained simply by permuting the states of one operation of the same type. For simplicity and generality, we focus in the following on gates that act in the 0-1 subspace of the qudit Hilbert space.

After the generation of the abstract sequence of rotations, the compiler simplifies the complexity of the problem by applying permutation gates to every $cRot$ and $pSwap$ gate in order to express the sequence in terms of only two entangling operations, $cRot_{1;0,1}$, $\qop{pSwap}_{0,1}$, and permutation gates. This way, every operation can be compiled at the expense of compiling efficiently only one $cRot$ and one $pSwap$. The two operations cannot, in general, be implemented directly on the hardware, therefore one further step is needed before concluding the synthesis.

Since $cRot_{1;0,1}$, $\qop{pSwap}_{0,1}$ are parametrized in $\theta$ and $\phi$, it would require to compile these default gates for every different value of the two variables.
The default gates chosen will be further decomposed into a sequence of local operations, that will encode the variables  $\theta$ and $\phi$ for the equatorial rotations, plus a static entangling gate that will provide the necessary entangling strength and that will act on a two-level subspace.

In this regard, the compiler will follow the decomposition of the controlled rotation $cRot_{1;0,1}$ (control on the level 1 of the first qudit, target the subspace 0-1 of the second qudit) as,

\begin{align}
    cRot_{1;0,1}(\theta,\phi)  = \ &[ I \otimes  R(\nicefrac{\pi}{2}, -\phi-\nicefrac{\pi}{2} )] \cdot \\ \notag
    & \qop{CEX}\cdot [I \otimes Z(\nicefrac{\theta}{2})] \cdot  \\  \notag
    & \qop{CEX}\cdot [I \otimes ( Z(\nicefrac{-\theta}{2}) \cdot  R(\nicefrac{-\pi}{2}, -\phi-\nicefrac{\pi}{2})  ), \\  \notag
\end{align}
where the decomposition uses the $\qop{CEX}{1;0,1}$ that will act on the 0,1 subspace of the second qudit, while the partial swap:
\begin{align}
    pSwap_{0,1}(\theta,\phi) = \ & 
    [H \otimes H  ]  \cdot \qop{CEX} \cdot [H \otimes H  ] \\  \notag
    & [I \otimes R(\nicefrac{\pi}{2}, -\phi-\nicefrac{\pi}{2})]  \cdot \qop{CEX} \cdot \\  \notag
    & [I \otimes Z(\nicefrac{\theta}{2})] \cdot \qop{CEX}  \cdot [I \otimes Z(\nicefrac{-\theta}{2})] \cdot \\  \notag
    & [I \otimes R(\nicefrac{-\pi}{2}, -\phi-\nicefrac{\pi}{2})] \cdot \\  \notag
    &  [H \otimes H  ]  \cdot \qop{CEX} \cdot [H \otimes H  ]
\end{align}
the $\qop{CEX}_{1;0,1}$ is used once again in the \qop{pSwap} operation.
\begin{example}
In order to better understand the intermediary compilation step, the operations involved, and the new encoding, let us consider the case where two qutrits are interacting. The controlled rotation on levels $\ket{7}$ and $\ket{8}$ appears in the decomposition, which correspond to $\ket{21}$ and $\ket{22}$ in the two-qutrit system. Since the method requires the controlled rotation to be applied to the 0-1 subspace, we have to permute the levels of the two-qutrit system. We add local rotations with angle $\pi$ and phase~$\nicefrac{\pi}{2}$ that will act as permutations and route $\ket{21}$ to $\ket{10}$ and $\ket{22}$ to $\ket{11}$.

     The decomposed sequence will be: 
     \begin{align*}
     cRot_{2;1,2}(\theta,\phi) = & (P_{0,2}\cdot P_{1,2} \otimes P_{0,1} \cdot P_{1,2})^\dagger \cdot\\
                                   &  cRot_{1;0,1}(\theta,\phi) \cdot\\
                                  &  (P_{0,2}\cdot P_{1,2} \otimes P_{0,1} \cdot P_{1,2}) ,
     \end{align*}
     where $P_{i,j}$ denotes a permutation of levels $i$ and $j$.

 Note that after executing the controlled-rotation, the sequence finishes with the uncomputation of the permutation sequence, in order to restore the original encoding for the subsequent operations.
 \end{example}

At this point the synthesis returns a sequence of local operations that will encode the angles of rotation and the permutation of the states, interleaved with the chosen \qop{CEX} gate.
This is the default choice made in the project, but the user could provide a different one that could act on a different subspace.
The compiler will now proceed with the last step: the compilation of the \qop{CEX} gate with an entangling gate input by the user and compatible with the target machine.

\subsection{Step 2: Layered Compilation}
\label{subsec:Ad-Hoc}
This section describes the second step of the compiler and the lowest abstraction level before getting to a physical implementation. The software component takes as input the target unitary to compile and the target entangling gate for the decomposition. The entangling gates could be a primitive entangling gate of the target quantum hardware, or it could be a higher-level abstract gate.
The software outputs a compiled sequence made of arbitrary local two-level rotations and the chosen gate. 

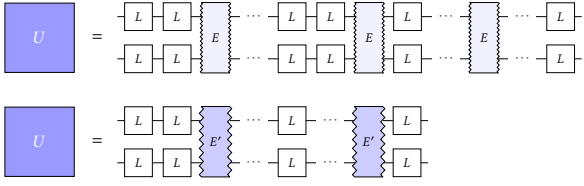
\begin{figure}[tp]
    \centering
    \resizebox{0.90\linewidth}{!}{\begin{tikzpicture}[font=\Large, dots/.style={draw=none, font=\large, yshift=0.2em}]
        \begin{scope}
            \node[fill=blue!40!white,text=white,draw=black, minimum size=2cm, font=\huge] (u) {$U$};
            \node[right=0.4cm of u, font=\huge] (eq) {$=$};
            \matrix[ampersand replacement=\&, matrix of math nodes, right=0.4cm of eq, row sep=0.4cm, column sep=0.3cm, nodes={minimum size=0.8cm, draw, fill=white}] (decompa) {
                L \& L \& |[draw=none, name=ent0a]|\phantom{L} \& |[dots]|\cdots \& L \& L \& |[draw=none, name=ent1a]|\phantom{L} \& L \& |[dots]|\cdots \& |[draw=none, name=ent2a]|\phantom{L} \& |[dots]|\cdots \& L \\
                L \& L \& |[draw=none, name=ent0b]|\phantom{L} \& |[dots]|\cdots \& L \& L \& |[draw=none, name=ent1b]|\phantom{L} \& L \& |[dots]|\cdots \& |[draw=none, name=ent2b]|\phantom{L} \& |[dots]|\cdots \& L \\
            };
            \begin{scope}[decoration={zigzag,segment length=0.4em, amplitude=.3mm}, line cap=round]
                \draw[fill=blue!5!white] (ent0a.north west) -- (ent0a.north east) decorate{-- (ent0b.south east)} -- (ent0b.south west) decorate{-- cycle};
                \draw[fill=blue!5!white] (ent1a.north west) -- (ent1a.north east) decorate{-- (ent1b.south east)} -- (ent1b.south west) decorate{-- cycle};
                \draw[fill=blue!5!white] (ent2a.north west) -- (ent2a.north east) decorate{-- (ent2b.south east)} -- (ent2b.south west) decorate{-- cycle};
            \end{scope}
            \path (ent0a) -- (ent0b) node[midway] {$E$};
            \path (ent1a) -- (ent1b) node[midway] {$E$};
            \path (ent2a) -- (ent2b) node[midway] {$E$};
            \begin{scope}[on background layer]
                \draw[thick, shorten >=-0.2cm, shorten <=-0.2cm] (decompa-1-1.west) -- (decompa-1-12.east);
                \draw[thick, shorten >=-0.2cm, shorten <=-0.2cm] (decompa-2-1.west) -- (decompa-2-12.east);
            \end{scope}
        \end{scope}
        \begin{scope}[yshift=-3cm]
            \node[fill=blue!40!white,text=white,draw=black, minimum size=2cm, font=\huge] (u) {$U$};
            \node[right=0.4cm of u, font=\huge] (eq) {$=$};
            \matrix[ampersand replacement=\&, matrix of math nodes, right=0.4cm of eq, row sep=0.4cm, column sep=0.3cm, nodes={minimum size=0.8cm, draw, fill=white}] (decompa) {
                L \& L \& |[draw=none, name=ent0a]|\phantom{L} \& |[dots]|\cdots \& L \& |[dots]|\cdots \& |[draw=none, name=ent1a]|\phantom{L} \& L \\
                L \& L \& |[draw=none, name=ent0b]|\phantom{L} \& |[dots]|\cdots \& L \& |[dots]|\cdots \& |[draw=none, name=ent1b]|\phantom{L} \& L  \\
            };
            \begin{scope}[decoration={zigzag,segment length=0.8em, amplitude=.6mm}, line cap=round]
                \draw[fill=blue!20!white] (ent0a.north west) -- (ent0a.north east) decorate{-- (ent0b.south east)} -- (ent0b.south west) decorate{-- cycle};
                \draw[fill=blue!20!white] (ent1a.north west) -- (ent1a.north east) decorate{-- (ent1b.south east)} -- (ent1b.south west) decorate{-- cycle};
            \end{scope}
            \path (ent0a) -- (ent0b) node[midway] {$E'$};
            \path (ent1a) -- (ent1b) node[midway] {$E'$};
            \begin{scope}[on background layer]
                \draw[thick, shorten >=-0.2cm, shorten <=-0.2cm] (decompa-1-1.west) -- (decompa-1-8.east);
                \draw[thick, shorten >=-0.2cm, shorten <=-0.2cm] (decompa-2-1.west) -- (decompa-2-8.east);
            \end{scope}
        \end{scope}
    \end{tikzpicture}}
    \vspace{-1em}
    \caption{
    Given a two-qudit unitary~$U$, compilation results depend on the structure and the noise inherent to the chosen gate $E$; new local ($L$) gates will match the gate $E$. 
    }
    \label{fig:problem-def}
    \vspace*{-2em}
\end{figure}

At the core of the second component are parametrized quantum circuits, that solved for different gates, lead to different decompositions. More precisely, as \autoref{fig:problem-def} shows, results present different depths and noise characteristics depending on the noise and power of the single entangling operation.

The section will continue with a brief overview of parametrized quantum circuits, followed by a detailed explanation on the construction of the problem representation and its resolution by the layered compiler.
\subsubsection{Parametrized Quantum Circuits (PQCs)}
PQCs consist of a series of quantum gates dependent on a set of continuous or discrete parameters $\theta_i$ that can be optimized. The particular initial choice of the sequence of gates is called an \emph{ansatz}. This circuit is then optimized, typically by a classical algorithm, to minimize a cost function that encodes the solution of interest. 

This approach can be used for a range of computational questions~\cite{NISQalgo}, like Hamiltonian ground-state energy estimation~\cite{Cerezo_2021} or compilation ~\cite{Martinez_2016}.
In the following we discuss the important design choices during the ansatz construction, its resolution and how the ansatz evolves during the optimization loop.

\subsubsection{Construction of Ansatz and Expressibility}
The first step consists of choosing the building blocks for our ansatz and then incrementally composing them in order to create the final circuit to be solved.
There are two types of operations involved in the quantum circuit: local operations and entangling operations.

The initial objective is to find a representation for single-qudit operations that is parametric and that achieves maximal expressibility~\cite{sim2019expressibility} with the minimal number of parameters. Here, expressibility refers to the circuit's ability to generate a large fraction of two-qudit unitaries. 
The proposed solution exploits theoretical results~\cite{Spengler2012} already present in literature. 
The single qudit lives in a $d$-dimensional $(d \geq 2)$ complex Hilbert space $\mathcal{H} = \mathbb{C}^d$ spanned by the orthonormal basis ${\ket{0}, \ldots, \ket{d-1}}$. Local qudit unitaries will be written as in~\cite{Spengler2012}, where elements of the special unitary group $\mathcal{S U}(d)$ are parametrized as:
\begin{align*}
    U =&\big[\textstyle\prod_{m=0}^{d-2} \big(\textstyle\prod_{n=m+1}^{d-1} \exp(i Z_{m, n} \lambda_{n, m}) \exp(i Y_{m, n} \lambda_{m, n})\big)\big] \cdot\\ 
    &\big[\textstyle\prod_{l=1}^{d-1} \exp (i Z_{l, d} \lambda_{l, l})\big]
\end{align*}
using $d^{2}-1$ real parameters $\left\{\lambda_{m, n}\right\}$ in the interval between $0$ and
$\pi$ for $m>n$, $\nicefrac{\pi}{2}$ for $m<n$, as well as $2\pi$ for $m=n$.

On this space, $Z$ is a diagonal trace-less operator, a generalized form of the Pauli $\sigma_{z}$ applied to the subspace spanned by~$\ket{m}$ and~$\ket{n}$, while $Y$ is the generalized form of the Pauli $\sigma_{y}$ applied to the subspace on the subspace spanned by~$\ket{m}$ and~$\ket{n}$. 
\begin{align*}
    Z_{m, n} &=\ket{m}\bra{m}-\ket{n}\bra{n}      &\text{for } 0 \leq m<n \leq d-1 \\
    Y_{m, n} &= -i\ket{m}\bra{n}+i\ket{n}\bra{m}  &\text{for } 0 \leq m<n \leq d-1
\end{align*}

This formulation comes with the promise of being expressible for the single qudits and it has the minimum number of parameters required for representing the special unitary group. 

The second objective is an efficient use of user-specified entangling gates in the ansatz compilation. The usage of multiple species of qudit entangling gates inside the same ansatz is future research.
Although the user can specify the preferred choice for the entangling primitive, here, we focus on using a single entangling gate, choosing between two gates commonly used in the trapped-ion platform: the Molmer-Sorenson~(MS) gate in \autoref{eq:ms}~\cite{ringbauer2021universal} and the generalized light-shift~(LS) in \autoref{eq:ls}, a genuine qudit entangling gate~\cite{Sawyer2021,nativequdit}.
\begin{align}
    MS(\theta) &= e^{-i\frac{\theta}{4} \cdot (I \otimes I + \sigma_{x_{01}} \otimes \sigma_{x_{01}})} \label{eq:ms} \\
    LS(\theta) &= e^{-i\theta\cdot \sum_{i=0}^{d-1} \ket{ii}\bra{ii}} \label{eq:ls}
\end{align} 
These are both well-established operations in quantum computing hardware with well-characterized noise characteristics. Moreover, they operate in a distinct way (the light-shift acts on phases, the MS acts on populations) with different entangling power~\cite{nativequdit} when applied to qudit systems.

\begin{figure}[tp]
    \centering
    \scalebox{0.75}{\begin{tikzpicture}[font=\Large, dots/.style={draw=none, font=\large, yshift=0.2em}]
        \begin{scope}
            \matrix[matrix of math nodes, row sep=0.4cm, column sep=0.4cm, nodes={minimum size=0.8cm, draw, fill=white}, ampersand replacement=\&] (decompa) {
                U_1 \& |[draw=none, name=ent0a]|\phantom{U_1} \& |[dots]|\cdots \& U_{f_1} \\
                U_2 \& |[draw=none, name=ent0b]|\phantom{U_2} \& |[dots]|\cdots \& U_{f_2} \\
            };
            \begin{scope}
                \draw[fill=white] (ent0a.north west) -- (ent0a.north east) decorate{-- (ent0b.south east)} -- (ent0b.south west) decorate{-- cycle};
            \end{scope}
            
            \path (ent0a) -- (ent0b) node[midway] {$E$};
            \node[left=0.4cm of decompa-1-1] {$Q_0$};
            \node[left=0.4cm of decompa-2-1] {$Q_1$};
            
            \begin{scope}[on background layer]
                \draw[shorten >=-0.2cm, shorten <=-0.2cm, transform canvas={yshift=-6pt}] (decompa-1-1.west) -- (decompa-1-1.east-|decompa.east);
                \draw[shorten >=-0.2cm, shorten <=-0.2cm, transform canvas={yshift=0}] (decompa-1-1.west) -- (decompa-1-1.east-|decompa.east);
                \draw[shorten >=-0.2cm, shorten <=-0.2cm, transform canvas={yshift=6pt}] (decompa-1-1.west) -- (decompa-1-1.east-|decompa.east);
    
                \draw[shorten >=-0.2cm, shorten <=-0.2cm, transform canvas={yshift=-6pt}] (decompa-2-1.west) -- (decompa-2-1.east-|decompa.east);
                \draw[shorten >=-0.2cm, shorten <=-0.2cm, transform canvas={yshift=0}] (decompa-2-1.west) -- (decompa-2-1.east-|decompa.east);
                \draw[shorten >=-0.2cm, shorten <=-0.2cm, transform canvas={yshift=6pt}] (decompa-2-1.west) -- (decompa-2-1.east-|decompa.east);
            \end{scope}
        \end{scope}
    \end{tikzpicture}}
	\caption{Example of an ansatz. Each layer contains an entangling gate, preceded by a generic unitary $U$ on each one of the qudits and followed by two more generic unitaries; in this way, every layer is universal.
	}
	\vspace*{-1.5em}
	\label{fig:layer}
\end{figure}
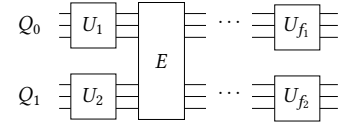

\begin{example}
    The compiler solves an ansatz like the one in \autoref{fig:layer}.--- 
    Consider a two-qudit entangling unitary~$U$. \autoref{fig:problem-def} illustrates two different compilation results:
    (1) The top result uses multiple pre-selected entangling gates with low entangling power but potentially less noise.
    (2) The bottom result uses two pre-selected entangling gates with high entangling power but more noise per gate.---
    The selection of the best result heavily depends on the performance of the gates, i.e.,~specifics of the underlying hardware. Both cases will always produce a correct result, as more entanglement can be built up with additional gates, and entanglement generation can also be reduced through judicious use of single qudit gates between entangling layers.
\end{example}

\subsubsection{Solution of the Ansatz}
The optimization of the ansatz works similar to a variational quantum algorithm. In each step, the full circuit is simulated by multiplying the gate matrices for the chosen parameters. This is computationally feasible for two-qudit gates in dimensions of practical relevance given the capabilities of current and future quantum hardware. For multi-qudit gates, the applicability of this approach is expected to be limited. The resulting unitary matrix in each optimization step is then compared to the target matrix according to an objective function.
Here, the choice was the fidelity between the unitaries~\cite{Wang2008AnAQ}, for two-$d$-dimensional systems:
\begin{align*}
    \textit{Fidelity}(A,B) = \frac{1}{d^2} \Tr{\langle A^\dagger \; , B \; \rangle}
\end{align*}
For the purpose of this work, the fidelity has several desirable properties that make it desirable for physical applications, compared to other commonly used objective functions for matrix optimization.

The optimization loop is performed by iteratively simulating the ansatz and verifying the fidelity achieved by the current solution of optimization algorithm. 

\subsubsection{Optimizer}
The used optimization method is the dual annealing method, a combination of classical and fast simulated annealing, coupled with the L-BFGS~\cite{Liu1989} algorithm on boxed constraints, which is a local search method applied on the neighborhood of the solutions found by the global method in the high dimensional landscape of the problem to optimize. We refer the reader to Ref.~\cite{2020SciPy-NMeth} for details on this algorithm. Alternatively, the use of gradient-based methods is left for future work, due to the unclear trainability of qudit PQCs.

\subsubsection{Binary Search}
Although the cost of a circuit depends on its depth and gates structures, the final resulting circuit should have the least number of layers that can compile with the desired fidelity the entangling gate. 
The maximum number of layers for the search is heuristically chosen as $2 \cdot d^2$. This number was found to be sufficient for generating maximally entangled qudit states from the least powerful operation, acting only on two fixed physical levels.
The compiler uses a binary search and the number of layers is decreased every time the gate can be decomposed for a certain number of layers under a time period heuristically chosen, otherwise if the target fidelity is not reached or the optimizer does not converge before the end of a timer, the number of layers is increased.

\section{Case Studies}
\label{sec:case-studies}
The considerations made so far in the paper already showed that the problem of entanglement compilation in high-dimensional systems is complex. The solution proposed above is supposed to provide a step forward in the field of compilation for qudit systems.
The implementation is available freely under the MIT licence at \href{https://github.com/cda-tum/qudit-entanglement-compilation}{\emph{github.com/cda-tum/qudit-entanglement-compilation}} as part of the Munich Quantum Toolkit (MQT). It is completely written in Python~3.8, with exception of the external dependencies Numpy~\cite{harris2020array} and Scipy~\cite{2020SciPy-NMeth}. This section provides corresponding case studies, in order to demonstrate the feasibility of the workflow and its usefulness in compiling any multi-level entangling operation.
The use cases focus on the compilation of the controlled-sum gate $\qop{CSUM}$, a characteristic operation in qudit systems as discussed in the previous sections.

Although constituted by two compilation steps, the computational complexity of the workflow is dictated on a high level by three routines.
The first routine is the QR decomposition applied on the unitary to compile; the algorithm has quadratic complexity in the number of dimension of the two-qudit interaction with size $D$, therefore $\mathcal{O}(D^2)$. The result is a sequence of, in the worst case, $D^2$~operations. The second routine is the translation of every output operation in term of local operations and \qop{CEX}; this algorithm has complexity $\mathcal{O}(D^2)$, because linear in the number of output operations of the QR.
Finally, the last routine of the workflow is the solution by optimization of the ansatz, which has complexity of $\mathcal{O}(\log{}D)$. In fact, the last compilation step is a binary search, with every step of duration linear in the number of dimensions of the original unitary.

The evaluations were performed on a server running GNU/Linux using an AMD Ryzen 9 3950X and \SI{128}{\gibi\byte}~main~memory. 
The layered compilation step is performed under a time limit in hours heuristically chosen as $\nicefrac{d}{4}$ (e.g.,~\SI{4}{\hour} for a two-ququarts unitary), at each step in the binary search. 
The compiler has minimum target infidelity ($1-\textit{Fidelity}$) of $10^{-3}$. 
Several runs of the optimization algorithm could lead to better results and, beyond dimension 16, the computational time and the fidelity are affected.

The considered example follows the default strategies encoded in the compiler. Every controlled rotation and partial swap is transformed by local rotations into 
$cRot_{1;0,1}$ and for the partial swaps $pSwap_{0,1}$, while the automatic choice for generating entanglement is the controlled exchange~gates~ \qop{CEX}.
More precisely, $\qop{CEX}_{1;0,1}$ has control on the first level of the first qudit and the targets on the first two levels of the second qudit.

The designer can decide to follow a different path and impose a decomposition for which they have a specific implementation in terms of the entangling gate involved, or compile it directly for the machine. 

\begin{table}[tpb]
    \centering
    \caption{Results of the workflow on CSUM for dimension 9 and dimension 16}
    \label{table:results}
    \vspace{-1em}
    \resizebox{\linewidth}{!}{\begin{tabular}{l r r r r r r r r}
        \toprule
        System & Dim. & $\qop{cRot}$ & $\qop{pSwap}$ & $\qop{CEX}_{tot}$ & $\qop{MS}_{tot}$ & $\qop{LS}_{tot}$ & $(1-F)$ \\
        \midrule
         two qutrits & 9 & 44 & 24 & 184 & 1472 & 368 & $<10^{-4}$ \\ 
         two ququarts & 16 & 92 & 36 & 328 & 9184 & 10168 & $10^{-3} \sim 10^{-4}$ \\
         \bottomrule
    \end{tabular}}
    \vspace*{-1.5em}
\end{table}

In the studies, we considered the implementation of the \qop{CSUM} gate for qutrits and ququarts.
\autoref{table:results} shows the results. Between the near-term usable physical qudit dimensions~\cite{ringbauer2021universal}, the solution for ququarts proves an already difficult and useful design case.
The columns denote the dimension (\enquote{Dim.}), followed by the number of controlled rotations, and partial swaps, as well as the total number of \qop{CEX}, \qop{MS}, and \qop{LS} gates required for the decomposition, respectively.
The last column gives a bound on the achieved infidelity (i.e.,~$1-\textit{Fidelity}$).
In a first phase the synthesis of \qop{CSUM} for 9 dimensions (two qutrits) requires 20 controlled rotations and 6 partial swaps, while the synthesis on two ququarts (dimension~16) outputs 42 controlled rotations and 6 partial swaps. 
These decompositions are, by construction, not necessarily optimal. 
The controlled rotations are further decomposed into 2 \qop{CEX}s and the partial swap into 4 \qop{CEX}s, with the resulting sequence being general for all dimensions.

In order to show to the adaptability of the workflow, in the smaller case we compile the entangling gate in terms of two native gates used in the latest ion trap qudit processors~\cite{ringbauer2021universal}~\cite{nativequdit} and introduced before, the \qop{MS} and \qop{LS} gate.
The manual decomposition of the \qop{CEX} gate dedicated for the two-qutrit case by itself is non-trivial, even for an experienced quantum information specialists. 
The inherent difficulty arises not only from the amount of intuition needed, but also from the inability to easily generalize the single result found to higher dimensions.
The pen-and-paper optimized sequence for \qop{CEX} in dimension 9 is made of two \qop{LS} gates with angle $\pi$, while two \qop{MS} gates are also required for performing the same \qop{CEX} at any dimension, if the sequence is allowed to exploit auxiliary levels~\cite{ringbauer2021universal}.

The layered compiler performs a similar decomposition with the same number of $\pi$-\qop{LS} gates autonomously. 
Without auxiliary levels, the compiler outputs a sequence of 8 \qop{MS} gates. For both the compilation results, the infidelity reached is below~$10^{-4}$. 
In the case of two-ququarts, optimized sequences could not be achieved, while the layered compiler autonomously decomposes the \qop{CEX} with infidelity between $10^{-3}$ and~$10^{-4}$, at the cost of 28 \qop{MS} gates and in alternative of 31~\qop{LS} gates.
The number of gates required for a two-ququarts circuit gives a feeling of the increased complexity of the problem compared to two-qutrits.

The results are in line with the expected \emph{automation overhead} in a first automatic solution.
In fact, it was predictable that the used ansatz would ask for an over-parametrization~\cite{Larocca2021}, or using more layers than necessarily, which allows for a great simplification of the landscapes by eliminating spurious local minima.
Although the results are not expected be efficient to the point of practical applicability on a quantum system, this is only one component of a complete compiler for high-dimensional and mixed-dimensional systems; future developments will comprise optimization routines\cite{da2013global, Duncan_2020, Kissinger_2020, qufinite, mato2022adaptive}.

In summary, the results show the feasibility of the proposed workflow to deliver decompositions efficiently for any two-qudit unitary.
Further, the results provide evidence compilation of entangling operations for multi-level systems is not easy, especially compared to binary systems. This work investigates the first key step towards full automation for qudits.
The implementation is available under the MIT license at \href{https://github.com/cda-tum/qudit-entanglement-compilation}{\emph{github.com/cda-tum/qudit-entanglement-compilation}} as part of the Munich Quantum Toolkit (MQT).

\section{Conclusion}
\label{sec:conclusion}
The challenges for efficient and reliable application of entangling gates inside qudit circuits arise because of the complexity in understanding the amount of entanglement generated by an operation as well as the corresponding affected levels---commonly referred to as structure. 
Consequently, the operations are mostly considered as black-boxes with the question of whether it is possible to reach an implementation for a certain qudit platform and with a given gate-set. 
So far, the study of feasibility and the implementation of entangling operations for specific qudit technologies was performed by quantum information specialists manually, without the promise of re-utilizing a particular decomposition for a system certain number of dimension, to those of greater dimensionality.

In this paper, we introduced a complete workflow for compiling any two-qudit unitary into any target gate set. 
While the resulting gate sequences are typically not optimal in terms of gate count, the method is computationally efficient, since the only step involving numerical optimization can be pre-computed.
The case studies confirm the feasibility of the workflow as well, for the first time,  the automated results of compilation of generic entangling operations to any dimensionality. 
The implementation is available under the MIT license at \href{https://github.com/cda-tum/qudit-entanglement-compilation}{\emph{github.com/cda-tum/qudit-entanglement-compilation}} under the ensemble of the Munich Quantum Toolkit (MQT).
In the future, the proposed approach may be improved up on by utilizing auxiliary qudit levels, alternative ansatz designs, compression of synthesis results, and circuit optimization in post-processing.

\section*{Acknowledgments} %
This work has received funding from the European Union's Horizon 2020 research and innovation programme under the ERC Consolidator Grant (agreement No 101001318), the Marie Sk{\l}odowska-Curie Grant (agreement No 840450), and the NeQST Grant (agreement No 101080086). It is part of the Munich Quantum Valley, which is supported by the Bavarian state government with funds from the Hightech Agenda Bayern Plus and was partially supported by the BMK, BMDW, and the State of Upper Austria in the frame of the COMET program (managed by the FFG). 

\printbibliography
\end{document}